\documentclass{iau}
\usepackage{graphicx}
\usepackage{caption}

\title[Spin Alignment in Analogues of The Local Sheet] 
{Spin Alignment in Analogues of The Local Sheet}

\author[George James Conidis]   
{George J. Conidis$^1$}

\affiliation{$^1$Department of Physics and Astronomy, York University, Toronto, Ontario, M3J 1P3, Canada \\ email: {\tt gconidis@yorku.ca}} 

\pubyear{2014}
\volume{308}  
\pagerange{119--126}
\jname{The Zeldovich Universe: Genesis and Growth of the Cosmic Web}
\editors{R. van de Weygaert, S. Shandarin, E. Saar \& J. Einasto}
\begin{document}

\maketitle

\begin{abstract}

Tidal torque theory and simulations of large scale structure predict spin vectors of massive galaxies should be coplanar with sheets in the cosmic web.  Recently demonstrated, the giants (K$_{s}$ $\leq$ -22.5 mag) in the Local Volume beyond the Local Sheet have spin vectors directed close to the plane of the Local Supercluster, supporting the predictions of Tidal Torque Theory.  However, the giants in the Local Sheet encircling the Local Group display a distinctly different arrangement, suggesting that the mass asymmetry of the Local Group or its progenitor torqued them from their primordial spin directions.  To investigate the origin of the spin alignment of giants locally, analogues of the Local Sheet were identified in the SDSS DR9.  Similar to the Local Sheet, analogues have an interacting pair of disk galaxies isolated from the remaining sheet members.  Modified sheets in which there is no interacting pair of disk galaxies were identified as a control sample.  

Galaxies in face-on control sheets do not display axis ratios predominantly weighted toward low values, contrary to the expectation of tidal torque theory.  For face-on and edge-on sheets, the distribution of axis ratios for galaxies in analogues is distinct from that in controls with a confidence of 97.6 $\%$ $\&$ 96.9$\%$, respectively.  This corroborates the hypothesis that an interacting pair can affect spin directions of neighbouring galaxies.

\keywords{cosmology: large-scale structure of universe, galaxies: spiral, galaxies: evolution, galaxies: interactions, galaxies: fundamental parameters}
\end{abstract}

\firstsection 
\section{Introduction $\&$ Overview}

The relationship between a galaxy's spin (angular momentum) vector and its surrounding environment is of substantial importance to the understanding of galaxy formation and evolution. In the last century, simulations of large scale structure have produced a better understanding of spin alignment in the cosmic web (Faltenbacher et al. 2002; Cuesta et al. 2008;
Paz et al. 2008; Zhang et al. 2009; Wang et al. 2011; Codis et al. 2012; Trowland et al. 2013; Zhang et al. 2013; Tempel et al. 2013; Tempel $\&$ Libeskind 2013; Cen 2014; Dubois et al. 2014). Investigations of alignment in specific environments (voids, sheets, filaments, and nodes) have been carried out using the well established tidal torque theory (Hoyle, 1949; Peebles, P. J. E., 1969; Efstathiou, G., et al. 1979; White,S. D. M. 1984; Porciani, et al. 2002; Schafer, B. M. 2009).  Within a sheet, simulations predict that dark matter haloes above 10$^{12}$ M$_{\odot}$ should have spin vectors preferentially directed along the mid-plane of the sheet (Arag\'on-Calvo et al. 2007; Hahn et al. 2007; Codis et al. 2012; Trowland et al. 2013).  However, such a prediction is untestable unless the assumption is made that spin vectors for dark and baryonic matter are aligned (van den Bosch et al. 2002; Yoshida et al. 2003; Chen, Jing $\&$ Yoshikawa 2003;  Kazantzidis et al. 2004; Springel, White, $\&$ Hernquist 2004; Bailin et al. 2005; Berentzen $\&$ Shlosman 2006; Gustafsson et al. 2006; Adabi et al. 2009; Croft et al. 2009; Romano-Diaz et al. 2009; Bett et al. 2010; Sales et al. 2012).  There have been many observational studies to look for alignment of baryonic spin with  cosmic structure, but results have been contradictory.  While some researchers have claimed to see alignment (Trujillo et al. 2006; Lee $\&$ Erdogdu 2007; Paz et al. 2008; Jones et al. 2010), others have found that spin directions are random (Slosar $\&$ White 2009; Cervantes-Sodi et al. 2010).   In those studies which find evidence for alignment, there is a preference for a disk galaxy to have its spin vector directed along the midplane of its host sheet (Trujillo et al. 2006; Varela et al. 2012; Tempel et al. 2013; Tempel $\&$ Libeskind 2013).  However, it is important to realize that observational studies constrain spin directions using axis ratios and position angles only.  For face-on and edge-on galaxies, there are two possible solutions, and for inclined galaxies there are four.  The ambiguity reduces the likelihood of a positive detection of alignment.

Galaxies within a distance of 6 Mpc from the Milky Way are localized in a highly flattened configuration known as the Local Sheet which is distinct from the Local Supercluster (Peebles et al. 2001; Tully et al. 2008; Peebles $\&$ Nusser 2010;  McCall 2014).  The extent (diameter) of the Sheet is 10.4 Mpc and the thickness (2$\times$ vertical dispersion) is 0.47 Mpc. The Sheet hosts 14 giants (K$_{s}$ $\leq$ -22.5 mag ), of which 2 are in the Local Group and 12 beyond.  Those beyond are distributed around a ring with a radius of 3.75 Mpc  (the Council of Giants) whose centre is only 1.1 Mpc from the centroid of the Local Group.

\begin{figure}[h]
\centering
  \includegraphics[width=\linewidth]{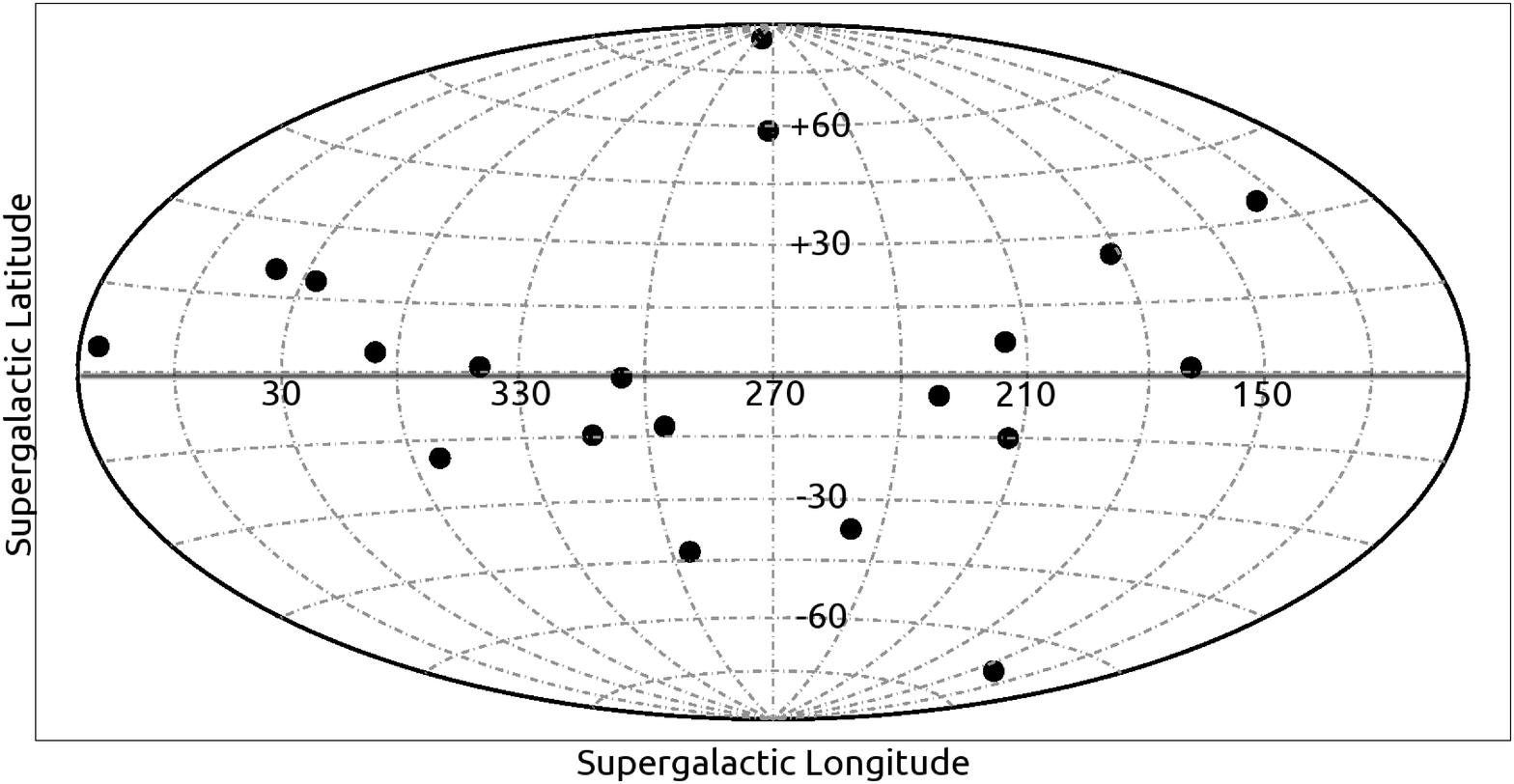}
  \caption{A hammer projection (area preserving) of spin vector directions of the giant galaxies in the Local Volume beyond the Local Sheet (black filled circles). The mean supergalactic latitude ($-1.8^{\circ}$) is shown as a horizontal blurred black line. The alignment of the giants with the supergalactic plane is in accordance with the prediction of tidal torque theory.}
  \label{fig:LVGs_SpinDist}
\end{figure}

McCall (2014) analyzed the distribution of spin vectors of the Local Sheet giants with respect to the giants beyond the Local Sheet in the Local Volume ($<$ 10 Mpc). The Local Sheet giants have a distinct spin distribution from the giants beyond, with vectors arranged around a small circle on the celestial sphere. However, the spin vectors for giants beyond the Local Sheet are arranged around a great circle on the sky which is closely aligned with the supergalactic plane (see Figure \ref{fig:LVGs_SpinDist}). The pole of this spin distribution, excluding the three outliers near the poles, is 23$^{\circ} \pm 13^{\circ}$ from the north galactic pole resulting in a sinusoidal trend in latitude. The mean supergalactic latitude of spin directions is only $-1.8^{\circ} \pm 20.6^{\circ}$. Indeed, coplanar spin vectors for galaxies in the luminosity range spanned is predicted by tidal torque theory.  

The distinct spin distribution observed for Local Sheet giants has been hypothesized to be the result of tidal torquing induced by the Local Group (McCall 2014). First of all, the Local Group contains 27$\%$ of the mass of the Sheet. By virtue of the pair of interacting giants which dominate it, its mass distribution is highly asymmetrical.  Also, its location is such that surrounding giants are at comparable distances; the radial dispersion is only $\pm 0.8$ Mpc.  Thus, the Local Group is the most likely source of torquing.

\section{Do interacting Pairs Influence Spin alignment in Local Sheet Analogues?}

To further our understanding of the effect of an interacting pair of galaxies on the spin alignments of galaxies in sheets, analogues of the Local Sheet were identified within the Sloan Digital Sky Survey Data Release 9 (Conidis $\&$ McCall, in preparation).  As controls, sheets were also identified in which the interacting pair is replaced with a single disk galaxy.  It was expected that if an interacting pair torqued fellow sheet members, the two samples would show a distinct difference.

\begin{figure}[h]
\centering
\begin{minipage}{.5\textwidth}
  \centering
  \includegraphics[width=\linewidth]{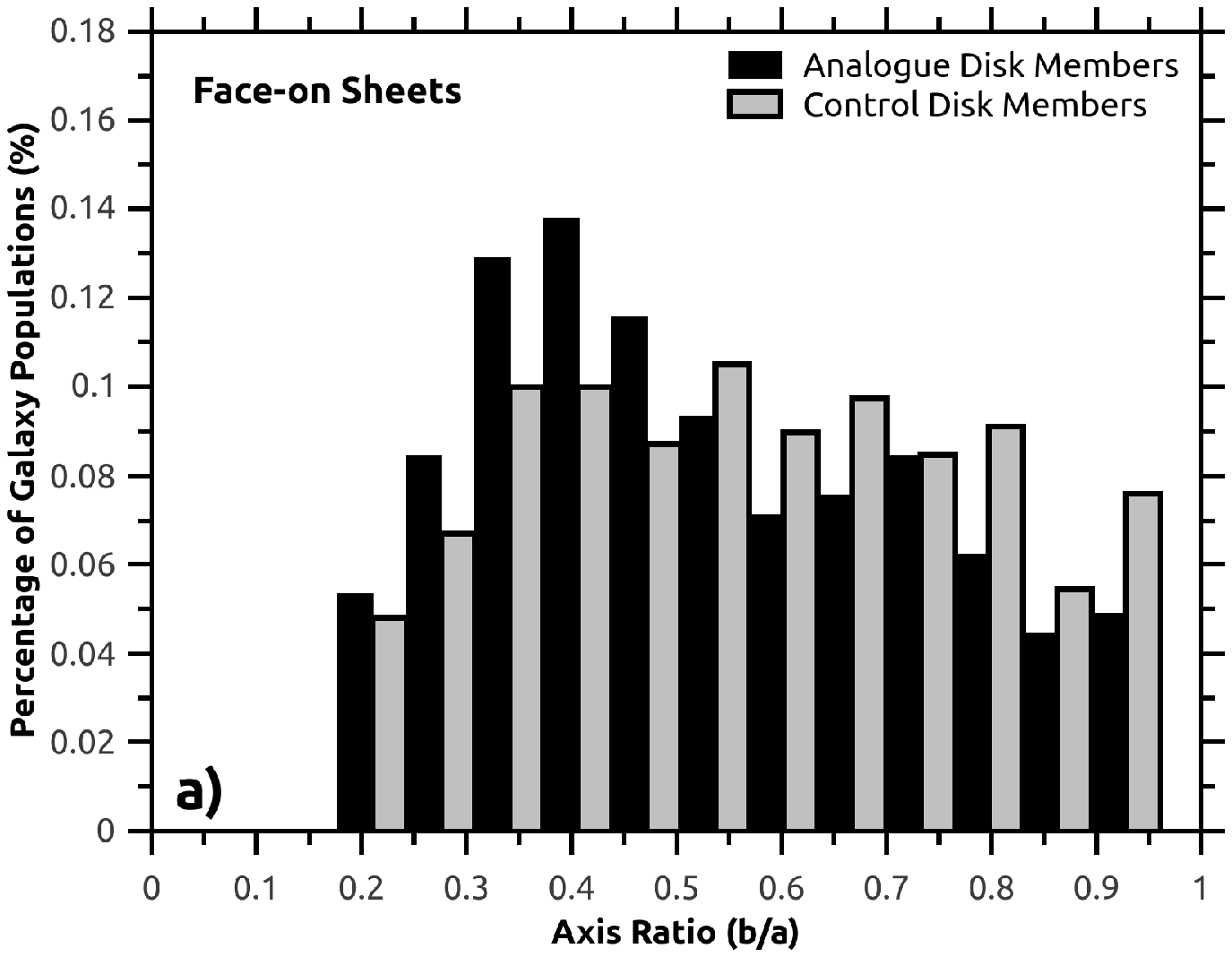}
  \label{fig:FaceOn_BAsHIST}
\end{minipage}%
\begin{minipage}{.5\textwidth}
  \centering
  \includegraphics[width=\linewidth]{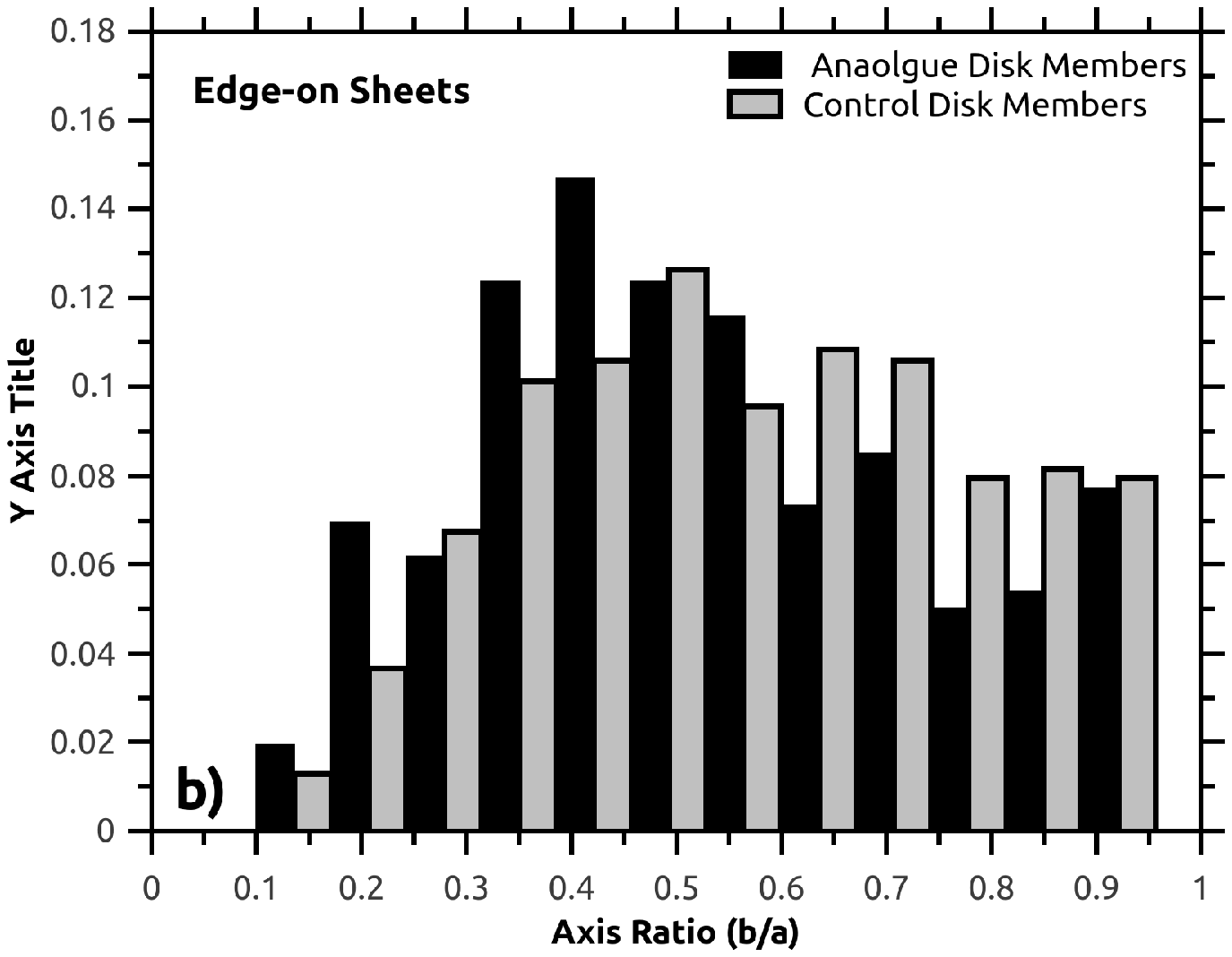}
 \label{fig:EdgeOn_BAsHIST}
\end{minipage}
\caption{Histograms of axis ratios of disks in analogue (black) and control (grey) sheets.  a) face-on sheets (tilt $< 44^{\circ}$).  b) edge-on sheets (tilt $> 70^{\circ}$).}
\label{fig:HISTs_BOTH}
\end{figure}

An assessment of spin distributions was carried out by examining axis ratios for disk galaxies in both analogue and control sheets.  Axis ratio distributions were derived for edge-on sheets (tilt $> 70^{\circ}$) and face-on sheets (tilt $< 44^{\circ}$). The control sheets were predicted to preserve their primordial spin organization, with spin vectors for members pointing along the midplanes of the sheets.  This prediction implied that axis ratios for galaxies in a face-on sheet would be clustered about low values. Figure \ref{fig:HISTs_BOTH}a (grey bars) is the observed axis ratio distribution for the disk galaxies in face-on control sheets. It does not cluster at low axis ratios as expected. It is found that the analogues disk galaxies hosted in a face-on sheet configuration tend to favour a low b/a value (Figure \ref{fig:HISTs_BOTH}a, black bars). In face-on analogue and control sheets, 56$\%$ and 46$\%$, respectively, of the galaxy population displays b/a $\leq$ 0.5. Furthermore the axis ratio distribution for face-on analogues and control have a 0.0029$\%$ and 10.1$\%$ chance, respectively, of being consistent with a random distribution. Clearly indicating the analogue and control distributions are distinct from each other. As well, the control's face-on distribution cannot be confidently distinguished from random. For edge-on analogue and control sheets (Figure \ref{fig:HISTs_BOTH}b), 54$\%$ and 45$\%$, respectively, of the galaxy population displays b/a $\leq$ 0.5. The axis ratio distributions for edge-on analogues and controls have a 0.025$\%$ and 0.57$\%$ chance, respectively, of being consistent with a random distribution. Thus, both distributions are found to be distinct from random.

\begin{figure}[h]
\centering
\begin{minipage}{.5\textwidth}
  \centering
  \includegraphics[width=\linewidth]{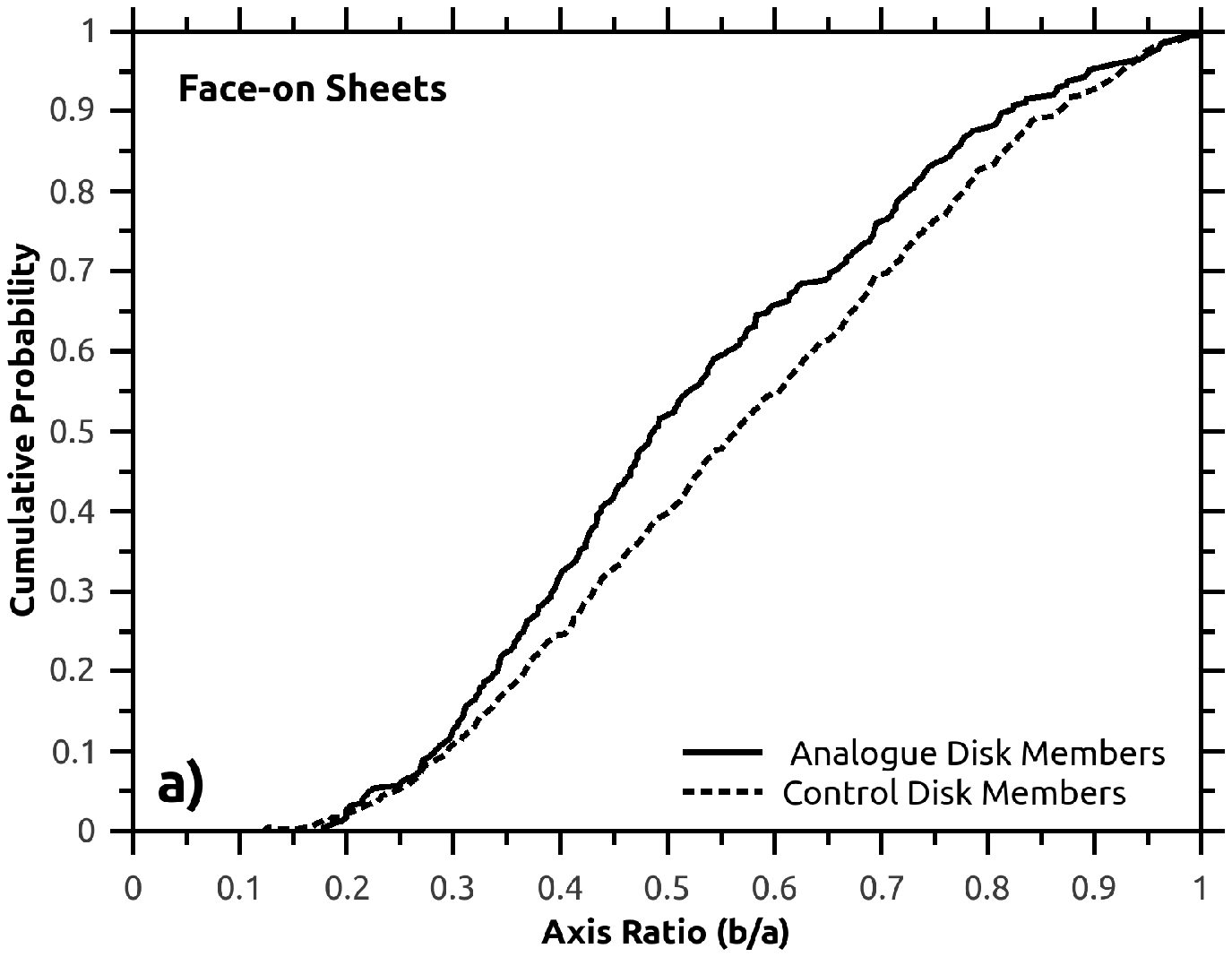}
  \label{fig:FACE_ON_CDFs}
\end{minipage}%
\begin{minipage}{.5\textwidth}
  \centering
  \includegraphics[width=\linewidth]{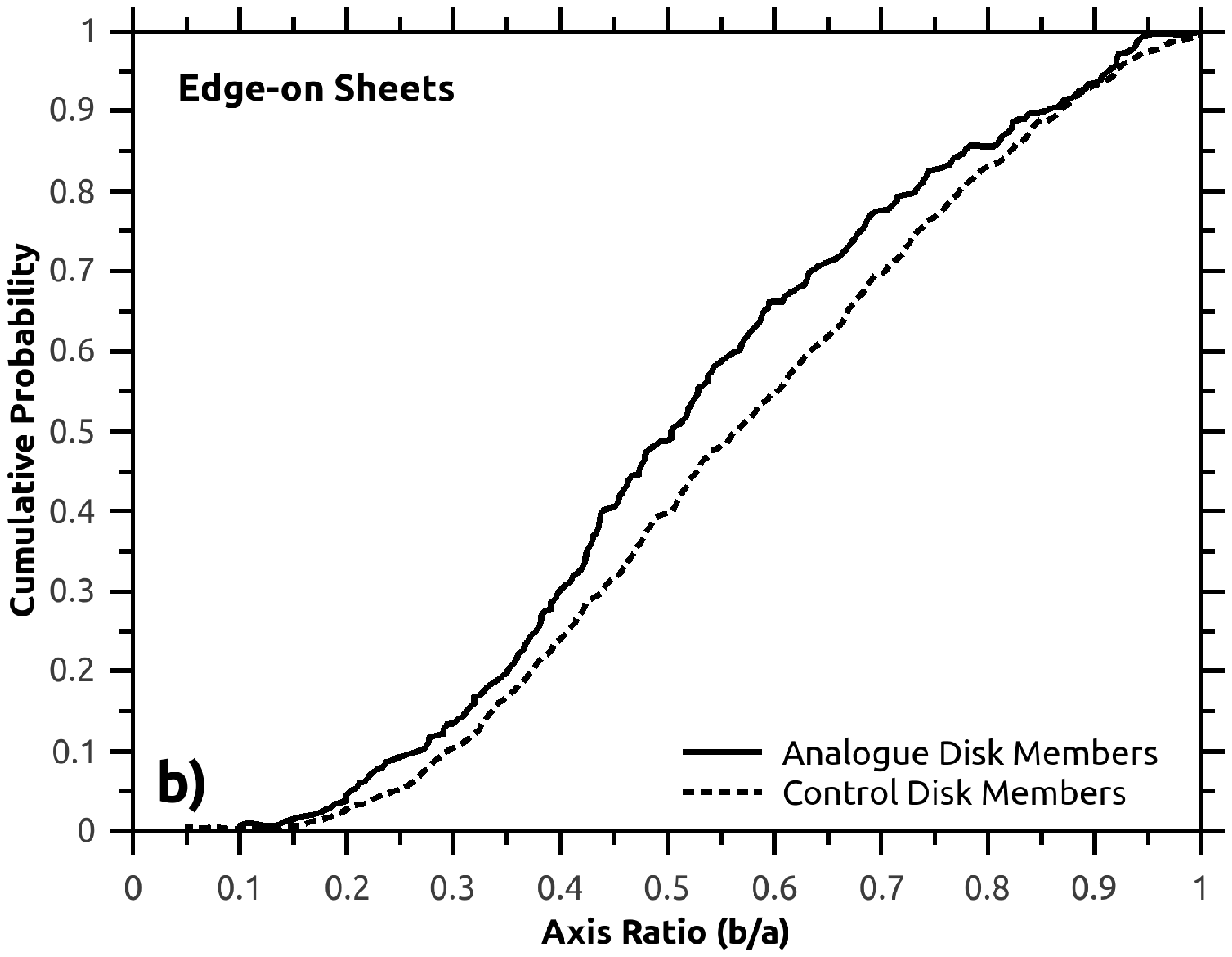}
 \label{fig:EDGE_ON_CDFs}
\end{minipage}
\caption{Cumulative distributions (CDs) of the axis ratios of disk galaxies in analogue (solid curve) and control (dashed curve) sheets. a) CDs for face-on sheets (tilt $< 44^{\circ}$). The Kuiper probability that these CDs were drawn from the same distribution is 2.4$\%$. b) Edge-on sheets (tilt $> 70^{\circ}$). The Kuiper probability that these CDs were drawn from the same distribution is 3.1$\%$.}
\label{fig:CDFs_BOTH}
\end{figure}

Cumulative distributions of the axis ratios of galaxies in analogue and control sheets are shown in Figure \ref{fig:CDFs_BOTH}.  The likelihood that the spin distributions for analogues and controls are drawn from the same probability distribution are 2.4$\%$ and 3.1$\%$, respectively, for face-on and edge-on orientations. This clearly shows that an interacting pair of disk galaxies embedded near the centre of a sheet torques the spins of surrounding members. 

\section{Implications and Future Works}

The distinct difference between the axis ratio distributions for analogue and control sheets, be they face-on or edge-on, validates the hypothesis that an interacting pair of galaxies changes the spin distribution of other members. This supports the conjecture that the Local Group was responsible for torquing the spins of galaxies in the Local Sheet out of alignment with the plane of the Local Supercluster.

Axis ratio distributions for galaxies in control sheets disagree with the predictions of tidal torque theory.  Especially, in face-on sheets, axis ratios do not cluster around low values and the distribution can be considered random within errors.  Surprisingly, the axis ratio distribution for face-on and edge-on analogues show a stronger tendency toward low axis ratios. As well, both these distributions are distinct from random providing additional evidence that a galaxy's orientation is environmentally dependant.

Analyses here are limited by the use of axis ratios as proxies for spin vectors.  For a given axis ratio, there are up to four spin directions possible. Currently, the author is engaged in determining unambiguously three-dimensional spin vectors for the disk galaxies in sheets, which will make possible more definitive statements about alignments (or lack thereof) with structure.

\section*{Acknowledgements}

I would like to thank Prof. Marshall L. McCall for his insightful discussions and helpful comments while conducting this research and producing this document.

\begin{discussion}

\discuss{Questioner\#1}{}
\discuss{Conidis}{}

\discuss{Questioner\#2}{}
\discuss{Conidis}{}

\discuss{Questioner\#3}{}
\discuss{Conidis}{}

\discuss{Questioner\#4}{}
\discuss{Conidis}{}
\end{discussion}

\end{document}